\documentclass[prb,twocolumn,showpacs,amsmath,amssymb,superscriptaddress]{revtex4-1}
\usepackage{dcolumn}
\usepackage{bm,graphicx}
\usepackage{etoolbox}
\usepackage{url}
\usepackage[colorlinks]{hyperref}
\usepackage{breakurl}
\usepackage{color}
\usepackage{amsmath}
\usepackage{xcolor}
\usepackage{soul}

\newcommand{\ee}{\varepsilon}
\newcommand{\kk}{\mathbf{k}}

\newcommand{\vo}{\varrho}

\newcommand{\s}{\sigma}

\newcommand{\om}{\omega}
\newcommand{\al}{\alpha}
\newcommand{\D}{\Delta}

\def\mathclap#1{\text{\hbox to 0pt{\hss$\mathsurround=0pt#1$\hss}}}

\begin{document}


\title{Probing intraband excitations in ZrTe$_5$: a high-pressure infrared and transport study}

\author{D.~Santos-Cottin}
\affiliation{Department of Physics, University of Fribourg, Chemin du Mus\'ee 3, 1700 Fribourg, Switzerland}

\author{M.~Padlewski}
\author{E.~Martino}
\affiliation{IPHYS, EPFL, CH-1015 Lausanne, Switzerland}

\author{S.~Ben David}
\author{F.~Le Mardel\'e}
\affiliation{Department of Physics, University of Fribourg, Chemin du Mus\'ee 3, 1700 Fribourg, Switzerland}

\author{M.~D.~Bachmann}
\affiliation{Max Planck Institute for Chemical Physics of Solids, N\"othnitzer Strasse 40, 01187 Dresden, Germany}
\author{C.~Putzke}
\author{P.~J.~W.~Moll}
\affiliation{Max Planck Institute for Chemical Physics of Solids, N\"othnitzer Strasse 40, 01187 Dresden, Germany}
\affiliation{Laboratory of Quantum Materials (QMAT), Institute of Materials (IMX), \'Ecole Polytechnique F\'ed\'erale de Lausanne (EPFL), 1015 Lausanne, Switzerland}

\author{R.~D.~Zhong}
\author{G.~D.~Gu}
\affiliation{Condensed Matter Physics and
   Materials Science Department, Brookhaven National Laboratory, Upton,
   New York 11973, USA}

\author{H.~Berger}
\affiliation{IPHYS, EPFL, CH-1015 Lausanne, Switzerland}

\author{M.~Orlita}
\affiliation{LNCMI, CNRS-UGA-UPS-INSA, 25, Avenue des Martyrs, 38042 Grenoble, France}
\affiliation{Institute of Physics, Charles University in Prague, 12116 Prague, Czech Republic}

\author{C.~C.~Homes}
\affiliation{Condensed Matter Physics and
   Materials Science Department, Brookhaven National Laboratory, Upton,
   New York 11973, USA}

\author{Z.~Rukelj}\email{zrukelj@phy.hr}
\affiliation{Department of Physics, University of Fribourg, Chemin du Mus\'ee 3, 1700 Fribourg, Switzerland}
\affiliation{Department of Physics, Faculty of Science, University of Zagreb, Bijeni\v{c}ka 32, 10000 Zagreb, Croatia}

\author{Ana Akrap}\email{ana.akrap@unifr.ch}
\affiliation{Department of Physics, University of Fribourg, Chemin du Mus\'ee 3, 1700 Fribourg, Switzerland}

\date{\today}

\begin{abstract}
Zirconium pentatetelluride, ZrTe$_5$, shows remarkable sensitivity to hydrostatic pressure. In this work we address the high-pressure transport and optical properties of this compound, on samples grown by flux and charge vapor transport.
The high-pressure resistivity is measured up to 2 GPa, and the infrared transmission up to 9 GPa. The $dc$ conductivity anisotropy is determined using a microstructured sample.
Together, the transport and optical measurements allow us to discern band parameters with and without the hydrostatic pressure, in particular the Fermi level, and the effective mass in the less conducting, out-of-plane direction. The results are interpreted within a simple two-band model characterized by a Dirac-like, linear in-plane band dispersion, and a parabolic out-of-plane dispersion. 
\end{abstract}

\maketitle

\section{Introduction}
Zirconium pentatelluride, ZrTe$_5$, is presently amongst the most investigated topological materials. This compound was studied in the context of a possible chiral anomaly,\cite{Li2016} a suggested 3D Dirac dispersion,\cite{Chen2015,Chen2015m} for being a potential weak\cite{Xiong2017,Moreschini2016,Li2016a,Wu2016} or strong topological insulator,\cite{Manzoni2016,Chen2017} as well as for its anomalous Hall effect linked to a putative Weyl dispersion.\cite{Liang2018}
The true ground state of ZrTe$_5$ is an unresolved question. This is linked to the very small energy scales characterizing the electronic band structure near the Fermi level.\cite{Xiong2017,Martino2019} 
Moreover, the samples seem to be significantly influenced by their preparation method.\cite{Shahi2018}

\begin{figure}[h!]
    \includegraphics[trim = 0mm 0mm 0mm 0mm, clip=true, width=0.8\linewidth]{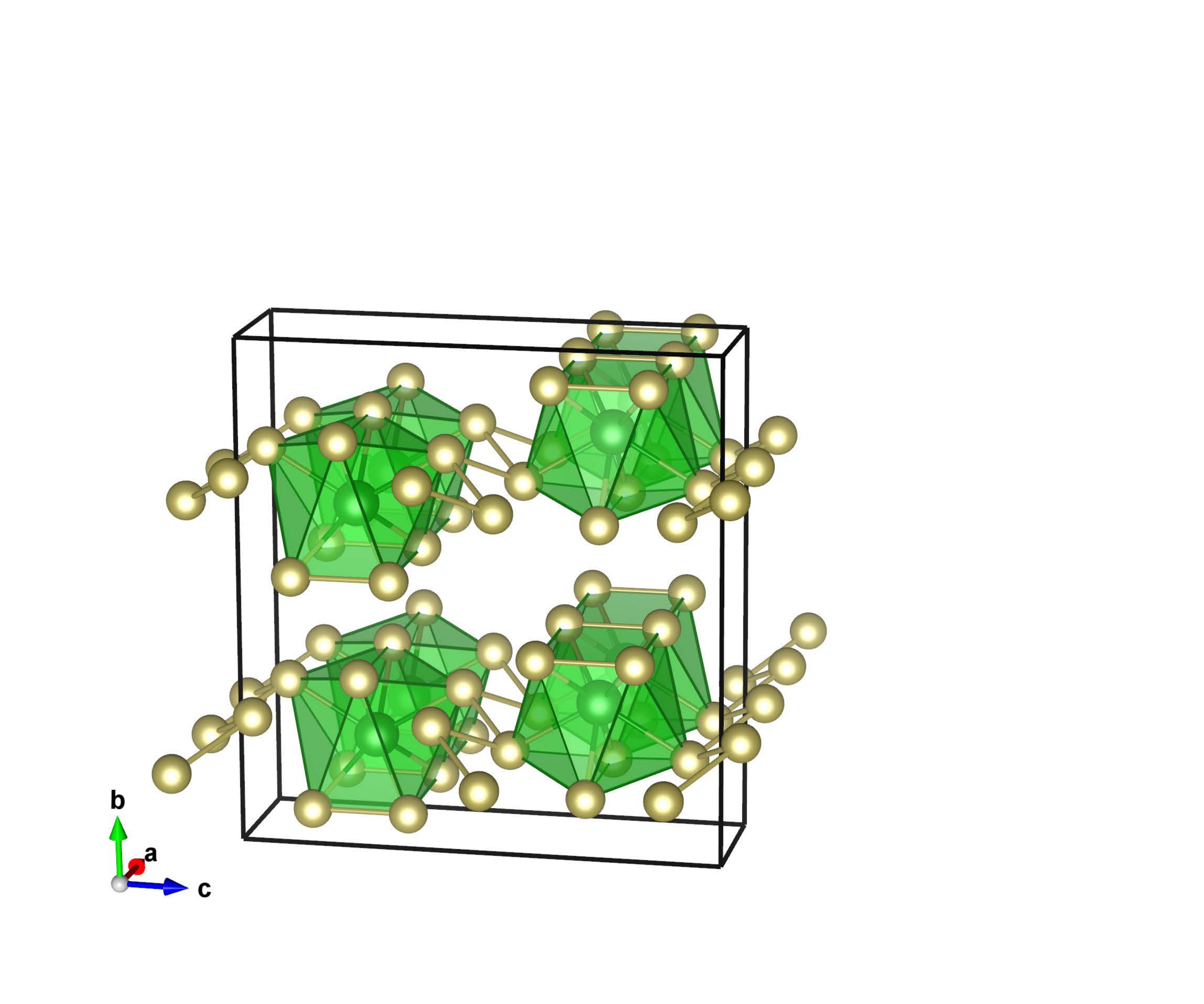}
\caption{\label{Fig0}Orthorhombic structure of ZrTe$_5$ in which zirconium atoms (green balls) are surrounded by tellurium atoms (yellow balls). The unit cell is shown by a solid line. The $b$ axis points between the planes, the zirconium chains run along $a$ axis.\cite{vesta}
}
\end{figure}

High pressure can be an excellent probe in a layered system such as ZrTe$_5$. Pressure changes the atomic orbital overlaps, in turn modifying the band structure. High-pressure behavior can often give a glimpse into the material's normal state properties.
ZrTe$_5$ is orthorhombic, as seen in Fig.~\ref{Fig0}, with its most conducting direction---$a$ axis---running along the zirconium chains. The layers are stacked along the least conducting, $b$ axis. Both the conduction and valence bands are based upon the tellurium $p$ orbitals. 
ZrTe$_5$ shows a remarkable sensitivity of the transport properties to hydrostatic pressure\cite{Zhou2016} and strain.\cite{Stillwell1989} Relatedly, thinning or exfoliating the crystals down to sub-micron thickness,  leads to large resistivity changes.\cite{Lu2017}
This sensitivity to lattice distortion is amplified by the small energy scales that characterize ZrTe$_5$.
The band gap is finite but very small, $2\Delta =6$~meV, and the carrier concentration can be made as low as $n \sim 10^{16}$ cm$^{-3}$, resulting in a significantly reduced Fermi surface.\cite{Martino2019} Very small carrier density means that a small magnetic field of $\sim 2$~T is sufficient to take the system into quantum limit, when all the carriers are confined to the lowest Landau level.\cite{Chen2015m}
Under high pressure ZrTe$_5$ becomes superconducting,\cite{Zhou2016} which underlines the importance of understanding the normal high-pressure state from which the superconductivity arises.

\begin{figure*}[th]
    \includegraphics[trim = 0mm 0mm 0mm 0mm, clip=true, width=1.\linewidth]{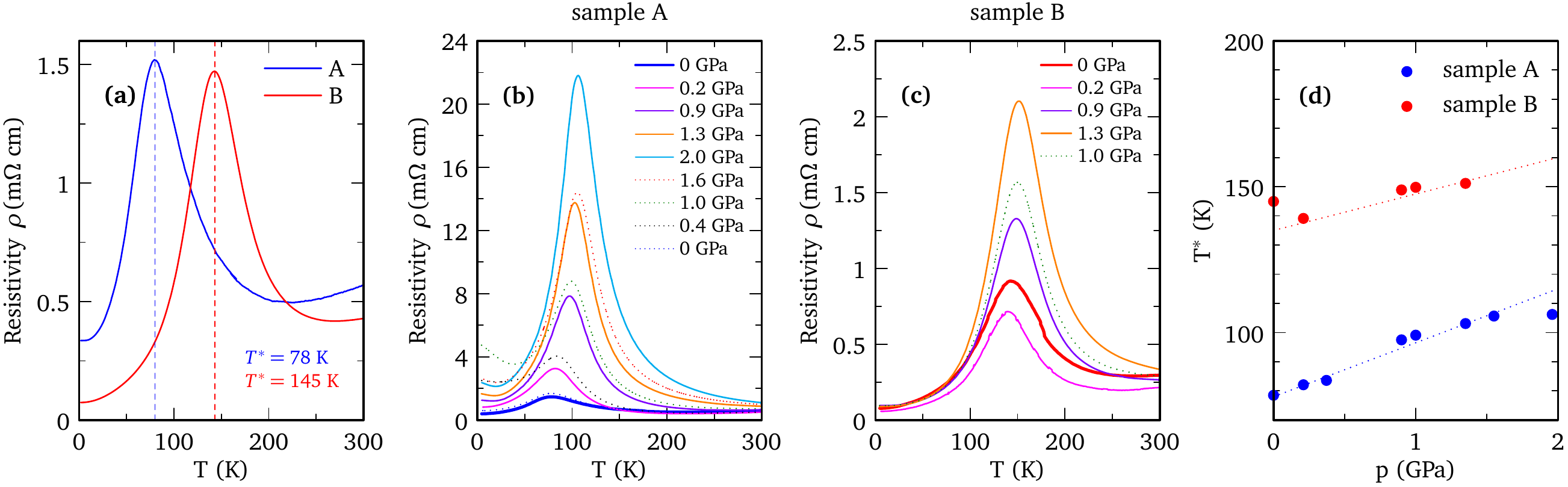}
\caption{\label{Fig1}(a) Ambient pressure resistivity of ZrTe$_5$, measured along $a$ axis, is shown as a function of temperature for both flux- and CVT-grown sample, sample A and B respectively. Dashed vertical lines denote the resistivity peak temperature, $T^*$. (b) Resistivity measured under several high pressures for (b) sample A and (c) for sample B. Full lines show resistivity for increasing the pressure, while dotted lines are curves for releasing the pressure.
(d) Pressure dependence of the temperature  $T^*$ where resistivity has a peak. Dotted lines are guides to the eye.}
\end{figure*}

Such malleable properties---by pressure, temperature, strain, and magnetic field---are desirable for many material applications. This is yet more interesting due to a simple, two-band nature of ZrTe$_5$ at low energies. However, we need to be certain to have a good understanding of the low-energy band structure.

Previously we have established an effective two-band model at low energies,\cite{Martino2019} based mostly on the interband optical excitations. In this work, we test this effective model using high pressure as a handle on the electronic properties. The high sensitivity of ZrTe$_5$ to pressure allows us to address the intraband, Drude response.
Through the high-pressure transport measurements, ambient pressure anisotropy study, and high-pressure optical transmission, we obtain valuable insight into the charge dynamics in ZrTe$_5$ at very low energies. 
Specifically, we show that the effective mass $m^*$ has a strong pressure dependence, and we explain the behavior of the $dc$ conductivity, as a function of pressure and temperature. 

\section{Experimental}

Measurements were performed on samples synthesized by two different methods. One batch of samples was made by self-flux growth,\cite{Li2016} and another by chemical vapor transport (CVT).\cite{Levy1983} 
Throughout this paper, we refer to flux-grown samples as sample A, and CVT-grown samples as sample B, although the measurements have been performed on multiple crystals of each batch, and are therefore reproducible.
The low-temperature carrier concentration is around 20 times higher in the CVT-grown sample than in the flux-grown sample.\cite{Martino2019, Shahi2018} At low temperatures these carrier concentrations are $n_A\approx 3\cdot 10^{-16}$~cm$^{-3}$ and $n_B \approx 6\cdot 10^{-17}$~cm$^{-3}$. 

Electrical resistivity was measured under high pressure inside a piston cylinder cell produced by C\&T Factory. 7373 Daphne oil was used as a pressure-transmitting medium to ensure hydrostatic conditions. Pressure was determined from the changes in resistance and superconducting transition temperature of a Pb manometer next to the samples.\cite{Eiling1981}
Electrical contacts to the sample were made using graphite conductive paint to ensure no degradation due to chemical reaction with the sample.

Optical reflectance is measured at a near-normal angle of incidence, using FTIR spectroscopy, with {\em in situ} gold evaporation.\cite{Homes1993} At high energies, the phase was fixed by ellipsometry. We use Kramers-Kronig relations to obtain the frequency-dependent complex dielectric function $\epsilon(\omega)$, where $\omega$ is the incident photon frequency. 
Analysis of the optical spectra was performed using RefFIT software.\cite{Kuzmenko2005}

\begin{figure*}[!hbt]
    \includegraphics[trim = 0mm 0mm 0mm 0mm, clip=true, width=0.95\linewidth]{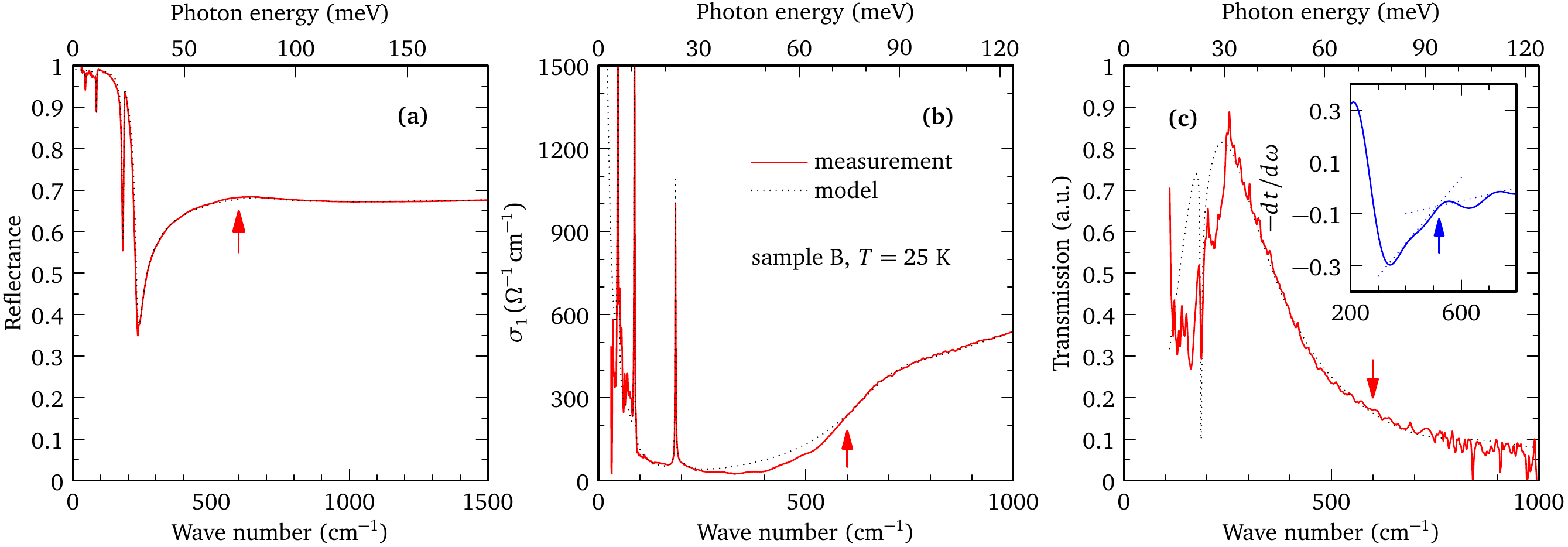}
\caption{\label{Fig2}Reflectivity (a), optical conductivity $\sigma_1$ (b) and optical transmission (c) are shown as a function of light frequency for sample B, at 25 K, with incident light polarized in the $a-c$ plane. In each panel, a Drude-Lorentz fit of reflectivity is shown in dotted lines, resulting in a calculated transmission and $\sigma_1$. Absorption onset or optical gap is marked by an arrow in each panel. It is visible as a hump in the reflectivity, a half-step in the optical conductivity $\sigma_1$. Inset in (c) shows the first derivative of transmission, in which a kink corresponds to the absorption onset. 
}
\end{figure*}

Transmission under high pressure was measured in a diamond anvil cell up to 9 GPa, at the base temperature of the setup, 25~K.
Synchrotron light passed through exfoliated micrometer-thin flakes of single crystals. The infrared experiments were done at the SMIS infrared beamline of Soleil synchrotron. High pressure was applied in a membrane diamond anvil cell, with CsI as a pressure medium. The anvils are made of IIa diamonds with 500~$\mu$m culet diameter. Applied pressure was determined from ruby fluorescence. 

The focused ion beam (FIB) microfabrication to create samples from resistivity anisotropy measurement,  was conducted using Xe plasma Helios G4 and Ga ion Helios G3 FIB microscopes manufactured by FEI. From a well characterized single crystal, a rectangular lamella was extracted along the desired crystallographic direction. After transferring to a sapphire substrate and gold contacts deposition by RF-sputtering, the final microstructure was patterned in the desired shape and individual electrical contacts created by etching through the deposited gold layer.\cite{Moll2018}
 

Finally, first principle calculations of band structure were done using density functional theory (DFT) with the generalized gradient approximation (GGA)
using the full-potential linearized augmented plane-wave (FP-LAPW) method \cite{singh}
with local-orbital extensions \cite{singh91} in the WIEN2k implementation \cite{wien2k}, as detailed in Ref.~\onlinecite{Martino2019}.

\section{Results}

Experimental data consist of resistivity under pressure, ambient and high-pressure infrared transmission, and finally a study of resistivity anisotropy on microstructured samples. We present the results in the same order below.

\subsection{High-pressure resistivity}

Figure \ref{Fig1} shows the resistivity along the $a$ axis for two high-mobility single crystal samples, made by the two methods mentioned above. The carrier mobilities are $\mu_H^A = 0.45\times 10^{6}$ cm$^2/$(Vs), and $\mu_H^B = 0.1\times 10^{6}$ cm$^2/$(Vs), for samples A and B respectively.\cite{Martino2019}
Specifically, Fig.~\ref{Fig1}a shows the ambient pressure resistivity of sample A (flux-grown) and sample B (CVT-grown). In both cases, the resistivity has a strong peak at a temperature $T^*$. This is 78~K for sample A, and 145~K for sample B. The resistivity peak is related to the low-temperature carrier density, so that a lower $T^*$ corresponds to a lower carrier density $n$,  and hence a lower Fermi level $\ee_F$.\cite{Martino2019,Shahi2018}
In both sample A and sample B, we observe a clear $\rho(T) \propto T^2$ behavior at low temperatures.\cite{Martino2019}

The peak temperature, $T^*$, is linked to dramatic change in many transport quantities. Thermopower and Hall effect change sign at this temperature, and the carrier density $n$ has a local minimum  at $T^*$. 
The resistivity peak seems to be linked to a minimum in carrier density at $T^*$.
The resistivity maximum has been the most puzzling experimental observation on ZrTe$_5$ in the past, and was thought to originate from a possible CDW. Presently it is understood that it comes from a temperature-induced shift of the chemical potential, with a concomitant crossover from low-temperature electron-dominated, to high-temperature hole-dominated conduction. 
Indeed, this link between $T^*$ and carrier density can be demonstrated even quantitatively.
The chemical potential shift also results in a $T^2$ behavior of the resistivity.\cite{Martino2019}

Figure \ref{Fig1}b shows the resistivity in the flux-grown sample A  taken at many different pressures, both while increasing the pressure (solid lines), and while releasing the pressure (dotted lines).
Surprisingly, the resistivity in sample A becomes 20 times higher under 2~GPa than at ambient pressure. 
This large effect is similar to the increase of resistivity in magnetic field.\cite{Tritt1999}
The resistivity peak strongly shifts in temperature, from 78~K at ambient pressure,  to 110~K at the highest achievable pressure, 2 GPa.
At higher pressures and in particular upon decreasing the pressure, the resistivity develops a low-temperature upturn. This upturn counterintuitively increases as the pressure decreases, showing a hysteretic behavior. Finally, the upturn disappears when pressure is completely removed. 

\begin{figure*}[ht]
    \includegraphics[trim = 0mm 0mm 0mm 0mm, clip=true, width=0.95\linewidth]{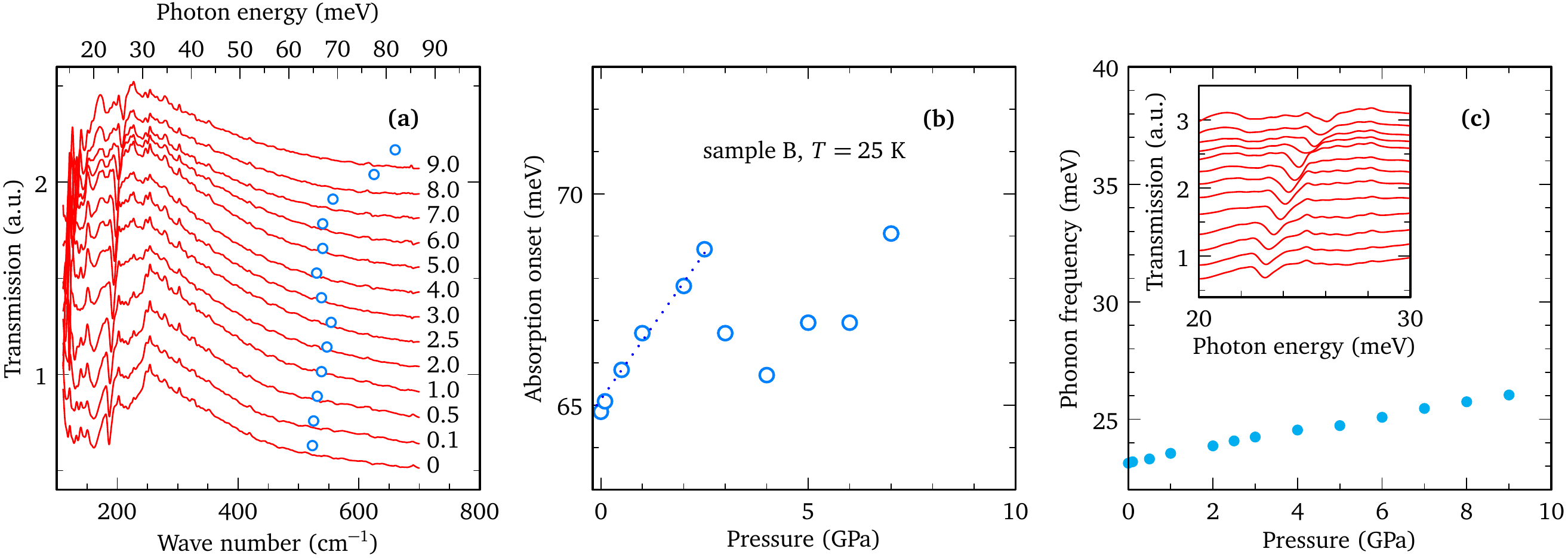}
\caption{\label{Fig3}(a) Infrared transmission at 25 K for sample B, taken at a series of pressures up to 9~GPa. Pressure values are in GPa, next to each transmission curve. Blue circles mark the position of a kink in the first derivative, which is taken as an approximate optical gap. (b) Extracted approximate value of the optical gap, or Pauli blocking edge, as a function of pressure. Dotted line serves to extract the parameter $\alpha$, as described in the text. 
(c) Pressure dependence of an infrared phonon at 23~meV (185~cm$^{-1}$) for ambient pressure.}
\end{figure*}

Similarly, Fig.~\ref{Fig1}c shows the resistivity in the CVT-grown sample B. Here too the peak temperature $T^*$ overall increases with pressure. While the absolute value of the resistivity increases under pressure, this effect is now much smaller than in sample A. Moreover, no upturns can be seen at low temperature.
These differences between samples A and B indicate that the chemistry of the sample growth strongly affects the high pressure behavior.
We speculate that the resistivity upturns may caused by the Kondo effect. A small amount of impurities might be clustered together inside the crystal by pressure. Alternatively,
these upturns may be related to a plastic deformation, possibly leading to a tunneling-like temperature dependence. 
The effect is much stronger in the flux-grown crystals, where all our high-pressure experiments show a resistivity upturn. While we do not have a clear explanation for the upturn, it is almost certain that it is an extrinsic effect, so we disregard it in the remaining discussion.

Figure \ref{Fig1}d shows the change of the resistivity peak $T^*$ as a function of pressure in samples A and B. In both cases the temperature $T^*$ increases under pressure, with an approximately linear, and rather large slope.
While our data only reaches 2~GPa, it is known from the literature that above 3 or 4 GPa, $T^*$ first starts to decrease and soon thereafter it disappears.\cite{Zhou2016}

At ambient pressure, the metallic resistivity well below $T^*$ is described by $\varrho=\varrho_0+AT^2$,\cite{Martino2019} with $A_A=0.1~\mu\Omega$cm/K$^{2}$ and $A_B=0.036~\mu\Omega$cm/K$^{2}$ for sample A and sample B, respectively. The coefficient $A$ is inversely proportional to $\ee_F$,\cite{Lin2015} indicating that the Fermi level in sample A is lower than in sample B.
The extracted $A$ seems to decrease under pressure, at least at very low pressures.
However, under higher pressure it becomes impossible to extract the prefactor $A$, due to the low-temperature upturns.


To summarize the measurements presented so far, we see that $T^*$ increases under pressure, and so does the absolute value of the resistivity at $T^*$. We notice a decrease in the $T^2$ resistivity prefactor, $A$. Finally, a strong upturn appears in the resistivity under pressure, but only in the flux-grown sample.

\subsection{Infrared transmission at ambient and high pressure}

Figure \ref{Fig2} shows ambient-pressure optical properties of sample B at 25 K. This is the temperature at which all of the high-pressure optical results will be discussed.
We elected to measure the sample B (CVT-grown sample) since its higher carrier density puts the optical gap in the accessible energy range under high pressure. In contrast, the Fermi level and the optical gap are much lower in the sample A, around 30~meV (200~cm$^{-1}$), and at the edge of our experimental window.
Figure \ref{Fig2} shows the ambient-pressure reflectivity (a), optical conductivity (b), and the ambient-pressure optical transmission (c) measured in a diamond anvil cell.
Transmission is more easily accessible under pressure than reflectivity. The other advantage is that the negative logarithm of transmission, $-\ln t$, for a thin, transparent sample behaves qualitatively similar to the real part of the optical conductivity, $\sigma_1$. This means that we may be able to extract Pauli blocking edge ($2\ee_F$) from infrared transmission. This quantity, $2\ee_F$, is equal to the energy of the onset of interband absorption.

\begin{figure*}[th]
    \includegraphics[trim = 0mm 0mm 0mm 0mm, clip=true, width=0.9\linewidth]{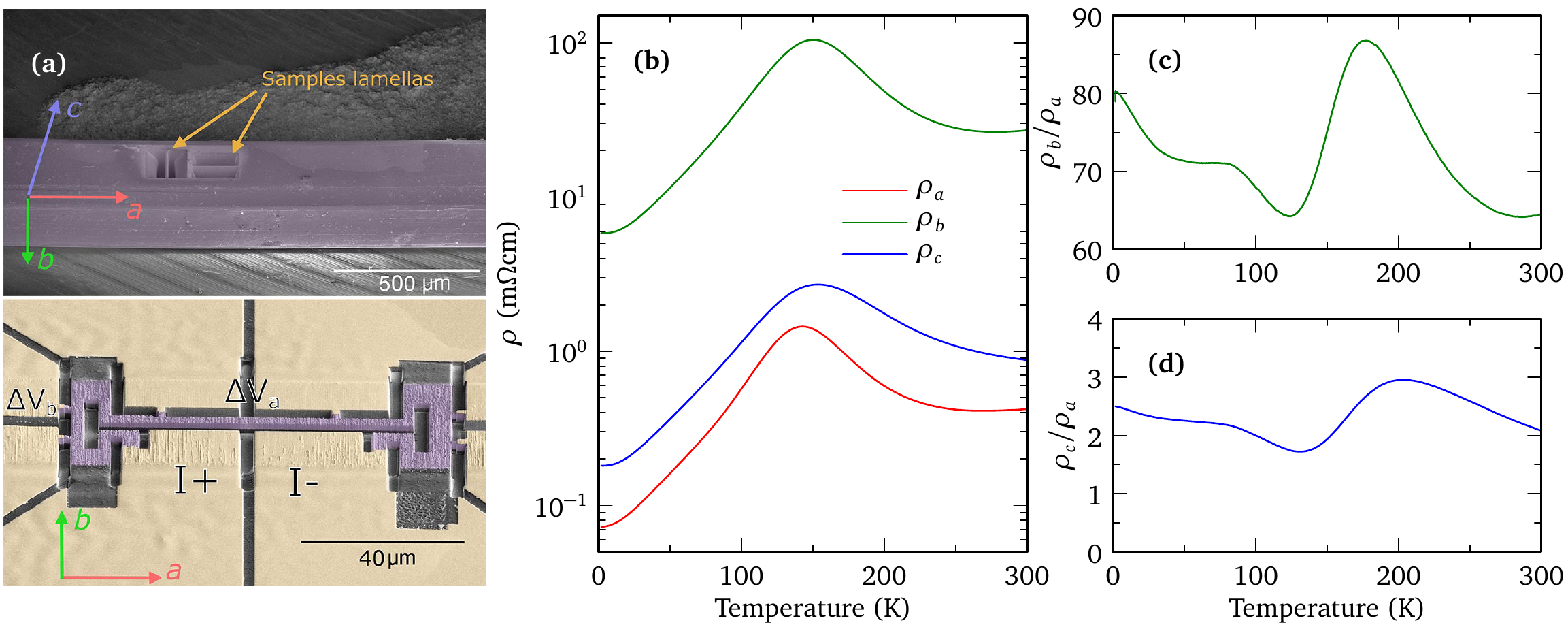}
\caption{\label{Fig4}Anisotropy of resistivity determined on a CVT-grown sample, prepared by focused ion beam technique. (a) Electron microscope picture of a microstructured sample prepared by focused ion beam technique. Top part shows the crystal from which two lamellas are shaped and cut. Bottom part shows a sample whose plane contains $a$ and $b$ axes. (b) Resistivity measured on sample B for all three crystal directions. Resistivity anisotropy (c) $\rho_b/\rho_a$ and (d) $\rho_c/\rho_a$ as a function of temperature.}
\end{figure*}

The main message of Fig.~\ref{Fig2} is that all the three quantities---reflectivity, optical conductivity, and transmission---are consistent.  
To show this, we apply a fit to the data using a Drude-Lorentz model\cite{Kuzmenko05} for the complex dielectric function:
\begin{equation}
  \tilde\epsilon(\omega)=\epsilon_\infty-\frac{\omega^2_{pD}}{\omega^2+i\omega/\tau_D}
          +\sum\frac{\Omega^2_j}{\omega^2_j-\omega^2-i \omega\gamma_j}.
\label{eq:eps}
\end{equation}
Here $\epsilon_\infty$ is the real part of the dielectric function at high
frequency, $\omega_{p,D}^2$ and $1/\tau_D$ are the square
of the plasma frequency and scattering rate for the delocalized (Drude) carriers,
respectively. For the Lorentz oscillators $\omega_j$, $\gamma_j$ and $\Omega_j$ are the position, width, and strength of the $j$th vibration or excitation.  
The same Drude-Lorentz model describes the reflectance and the optical conductivity $\sigma_1$, and it also agrees well with the measured transmission.

This self-consistency check is important when dealing with high-pressure results. While the reflectance and optical conductivity are linked by a Kramers-Kronig relation, it is not {\em a priori} evident that the optical transmission, separately measured inside a diamond anvil cell, with a pressure medium, will be free of extrinsic effects. The agreement between the experimental data and the Drude-Lorentz model also tells us that our experimental window for transmission is from 150 to 1000~cm$^{-1}$ (20 to 120~meV).

Finally, the data and the model show that there is no clear feature in transmission which can be associated with the Pauli blocking edge, or the onset of absorption, as illustrated in Fig.~\ref{Fig2}c. In the simplest approach, where $\sigma_1(\omega)$ around the Pauli edge resembles a step function, transmission $t(\omega)$ will have a kink in its first derivative, as illustrated in the inset of Fig.~\ref{Fig2}c. That is why, as an estimate of the optical gap (Pauli blocking) from our data, we take the position of the kink in $dt(\omega)/d\omega$.\\

Figure \ref{Fig3}a shows a series of transmission curves taken on sample B (CVT-grown) for a wide range of pressures reaching 9.0~GPa. For each pressure, blue circles mark the energy of the kink in the first derivative, and this value is taken as an estimated absorption onset or optical gap. Figure \ref{Fig3}b shows how this absorption edge evolves under pressure. From the ambient pressure up to 2.5~GPa it increases, followed by a more complex behavior up to the highest pressure reached.  This low-pressure increase of the absorption onset agrees with the increase of resistivity peak, which we will shortly show to be proportional to $\ee_F$. 
At pressures above 2.5~GPa, the behavior of the absorption onset is reversed and it drops  4~GPa, before increasing again. Similar nonmonotonic behavior is also observed in $T^*$ in function of pressure,\cite{Zhang2017} where a linear increase of $T^* (P)$ is followed by a decrease of $T^*$  around $1.5 - 2$~GPa. The nonmonotonic trend of $T^*$ suggests that the onset of absorption at higher pressures becomes influenced not only by the Fermi level, but also by changes in other parameters such as the band gap $\Delta$, or the Fermi velocities along all three axes. 

Figure \ref{Fig3}c shows the evolution of the only infrared-active phonon that is observed, and its ambient pressure frequency is at 185~cm$^{-1}$ (23~meV). 
This mode is likely the high-frequency $B_{3u}$ mode that is described by the Zr 
atom moving against Te atoms along the chain direction.\cite{vibratz}

The phonon hardening under pressure can be readily understood -- as the pressure makes the interatomic distances smaller, the vibrations become stiffer and move to higher frequencies.
The phonon behavior also confirms that our data is reliable down to 150 cm$^{-1}$ (20~meV).

The main experimental observation from the above high-pressure optical results is that the Fermi level $\ee_F$ linearly increases under pressure up to 2.5~GPa.
Throughout the whole pressure range, we see no dramatic change of low-energy transmission. This suggests that the low-energy band structure likely stays similar up to the highest pressures reached in this study.

\begin{figure*}[th]
    \includegraphics[trim = 0mm 0mm 0mm 0mm, clip=true, width=\linewidth]{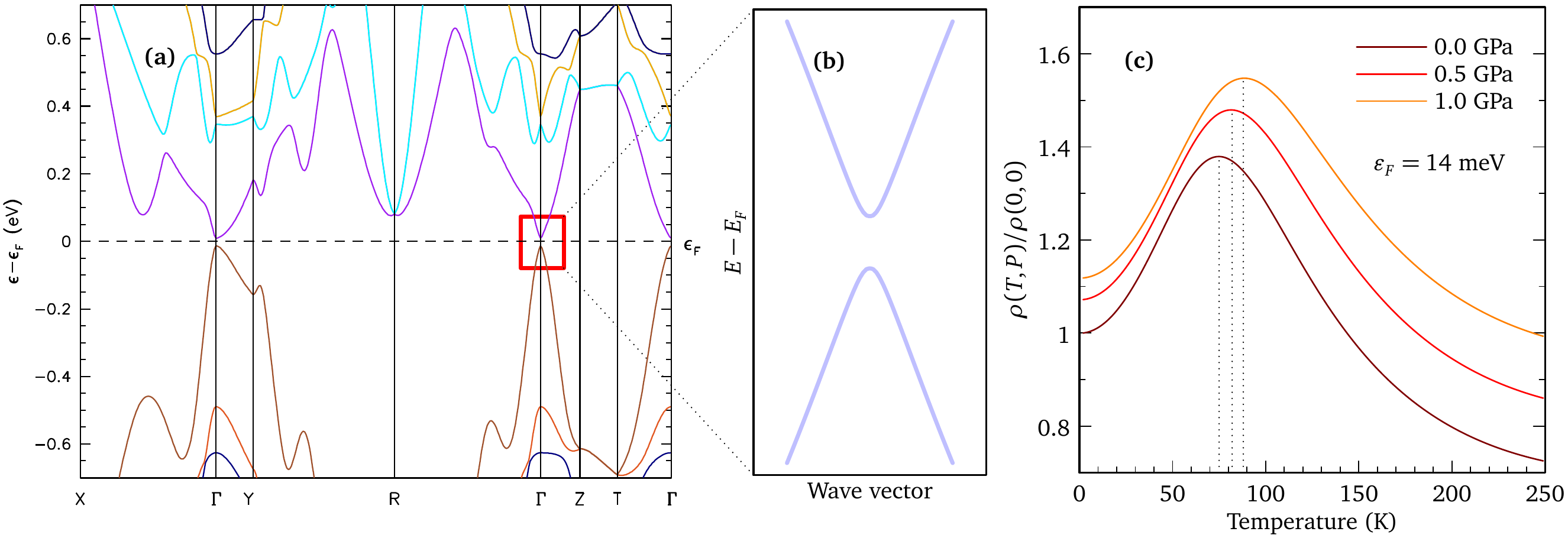}
\caption{\label{Fig5}(a) Electronic band structure determined by DFT, which considers spin-orbit coupling. (b) Schematic zoom around the $\Gamma$ point shows model bands $\epsilon_{1}$ and $\epsilon_{2}$.
(c) Numerically determined resistivity as a function of temperature, at three different pressures. Dotted vertical lines indicate the positions of the resistivity peak. 
}
\end{figure*}

\subsection{Anisotropy of conduction \\ in microstructured samples}

To complement the high-pressure data, we determined the anisotropy of the resistivity as a function of temperature, shown in Fig.~\ref{Fig4}. The measurement is performed on a microfabricated CVT-grown sample which allows to obtain a precise measurement of the resistivity for  the otherwise inaccessible out-of-plane $b$ direction. 
The measured resistivity ratio is high, $\rho_b/\rho_a \sim 80$ for the out-of-plane resistivity, and a modest $\rho_c/\rho_a \sim 2$ for the resistivity along the two in-plane axes. Interestingly, despite a large factor between the in-plane and out-of-plane resistivities, their temperature dependence is very similar.

The values of anisotropy change with temperature, especially around the resistivity maximum. This is due to a small, parasitic strain in the microfabricated devices. The temperature dependence of the resistivity anisotropy underlines the strong impact of pressure or strain on the low-energy conductivity.


\section{Theoretical model}

ZrTe$_5$ has a complex crystal structure with many atoms in the unit cell. The DFT-calculated band structure is shown in Fig.~\ref{Fig5}, where the spin-orbit coupling is taken into account.
While DFT gives the correct energy band dispersion at high energies,\cite{Martino2019} at the energies comparable to the very small, mili-electron-volt-sized spin-orbit gap, any {\em ab initio} technique will inevitably be unreliable. Instead, an effective model is necessary to account for all the low energy band features.

Most of the above experimental observations may be explained within a previously developed effective two-band model, with a linear energy dispersion in the $ac$ plane and a parabolic dispersion in the out-of-plane $b$ direction.\cite{Martino2019}
We can qualitatively understand the most important features of the $dc$ resistivity by considering only the intraband effects. The most distinct element---the resistivity peak at $T^*$---turns out to be a consequence of a strong chemical potential shift.

In the remainder of this Section, we refer to the axes $a$ and $c$ as $x$ and $y$. The least conducting $b$ direction is labeled $z$.

\subsection{Hamiltonian and its eigenvalues}

Let us start from a simple, two-band Hamiltonian with four free parameters which have been previously determined:\cite{Martino2019} 
\begin{equation}\label{jed}
H=\begin{pmatrix}
 \Delta + c k_z^2 & \hbar v_xk_x - i \hbar v_yk_y \\
 \hbar v_xk_x + i \hbar v_yk_y &  -\Delta - ck_z^2
\end{pmatrix} .
\end{equation}
The elements $v_{\al}$ are the Dirac velocities in the $x$ and $y$ direction (corresponding to $a$ and $c$ axis respectively), and $c = \hbar^2/2m^*$, where $m^*$ is the effective mass.
These parameters can be unambiguously determined by comparing the experimental data from the optical and transport measurement on $\rm{ZrTe_5}$,
and the predictions derived from the above Hamiltonian.
This Hamiltonian contains quasi-Dirac features in the vicinity of the  $\Gamma$   point in the Brillouin zone. It phenomenologically implements the energy band gap $2\D$ originating from the spin-orbit coupling, with an assumption of free-electron like
behaviour in the $z$ direction (along the $b$ axis).

The eigenvalues of the Hamiltonian (\ref{jed}) are
 \begin{equation}\label{dva}
\ee_{2,1 \kk} = \pm \sqrt{\hbar^2(v_xk_x)^2 + \hbar^2(v_yk_y)^2 + (\Delta + ck_z^2 )^2 }.
\end{equation}
These energies, $\ee_{1,\kk}$ and $\ee_{2,\kk}$, are symmetrical with respect to the middle of the bandgap, as illustrated in Fig.~\ref{Fig4}b.

\subsection{High-pressure behavior}
Our main assumption for the high-pressure analysis is that ZrTe$_5$ consists of layers which are weakly bound by van der Waals forces. In view of its layered structure, shown in Fig.~\ref{Fig0}, as well as the high resistivity anisotropy, this is a fair assumption. 

We examine the changes of the physical constants under the applied hydrostatic pressure $P$. Usually this situation is modeled only by {\it ab initio} calculations, but under a few reasonable presumptions, we can quantitatively determine several effects. There are 5 parameters that determine our system: Fermi velocities $v_x$ and $v_y$, effective mass $m^*$, energy gap $2\Delta$, and Fermi level $\ee_F$ at zero temperature. All of these parameters are susceptible to change as we apply hydrostatic pressure. However, not all of them change drastically under pressure. Since the Fermi velocities are connected to the slope of the electron
dispersions in the $x,y$ plane---a plane with strong interatomic forces---we suspect that the pressure will result in only a modest change of the velocities. Equivalently, one can say that the change of the $x,y$ plane area due to the pressure is insignificant. 
The effective mass $m^*$ describes the effects of interlayer forces which are much weaker than intralayer forces. Hence the main contribution to the volume decrease under pressure will come from the decreasing of the distance between the layers. Therefore, $m^*_P$ and $\ee_F^P$ are the two quantities that bear most of the pressure dependence.
While $m^*_P$ can be calculated, $\ee_F^P$ can be determined from our experiments.

We want to use the fact that the total carrier number $N$ remains constant under pressure. This is why we need to know how the density of states  and the volume depend on pressure.

We first calculate the single band density of states per unit volume using  the eigenvalues (\ref{dva}), and we obtain:
\begin{equation}\label{g2}
g(\ee) =  \frac{1}{\pi^2 \hbar^3} \frac{\sqrt{2m^*}}{v_xv_y}  \,\, \ee \sqrt{\ee- \Delta} = C \sqrt{m^*} \,\, \ee \sqrt{\ee- \Delta},
\end{equation}
where we introduce $C=\sqrt{2}/(\pi^2 \hbar^3 v_xv_y)$.

From the experimental dependence of the absorption onset on pressure below 2.5~GPa(Fig.~\ref{Fig3}b), we can deduce that 
\begin{equation}\label{eFp}
\ee_F^P = \ee_F(1+\al P) 
\end{equation}
where $\al = (1/\ee_F) \partial \ee_F / \partial P$. 
The coefficient $\alpha$ may be obtained from the Pauli edge shift in the optical experiment. Since the expression for the optical conductivity has the form 
\begin{eqnarray}\label{g5}
 \s_{xx}(\omega,0) = \frac{\s_0}{\pi} \frac{v_x}{v_y}\frac{\sqrt{m^*}}{\hbar}\sqrt{\hbar\omega - 2\D} \, \,\,\Theta(\hbar\omega - 2\ee_F),
\end{eqnarray}
we see that high pressure influences not only the amplitude of $\sigma_{xx}$ but also its onset, through the Pauli edge. The approximate value of the proportionality factor deduced from our experiment (Fig.~\ref{Fig3}b) is  $\alpha_B \approx 2.3 \cdot 10^{-10}$~Pa$^{-1}$, where we have employed the data for the sample B. 

Another way to obtain $\alpha$ is to look at the behavior of the resistivity anomaly under pressure, shown in Fig.~\ref{Fig1}d. 
The peak temperature $T^*_P$ is linear in pressure, $T^*_P = T^* (1 + \gamma P)$.
One can estimate $ \gamma_A \approx 2.4 \cdot 10^{-10}$~Pa$^{-1}$, and $\gamma_B \approx 0.86 \cdot 10^{-10}$~Pa$^{-1}$, from data on samples A and B respectively.

We can now use the relation between the resistivity maximum temperature $T^*$ and the Fermi energy, which is valid for our model:\cite{Rukelj2019}
\begin{equation}\label{g6}
T^*=const. +  (\ee_F/\Delta)  \cdot 17\,\rm{K}\
\end{equation}
From comparison between Eqs.~(\ref{eFp}) and (\ref{g6}), we can conclude $\alpha = \gamma (T^* \Delta)/(17\,\rm{K}\, \ee_F).$ 
This gives $\alpha_A = 2.3 \cdot 10^{-10}$~Pa$^{-1}$, where we use the parameters for sample A, $2\Delta= 6$~meV and $\ee_F=14$~meV, extracted from optical and magneto-optical measurements.\cite{Martino2019}

Similarly, the volume change under pressure can be written as $V_P = V (1-\beta P)$, where
the compressibility is defined as $\beta = -(1/V) \partial V / \partial P$. 
From the X-ray experiment under pressure\cite{Zhou2016} and using the above definition, we calculate the compressibility $\beta \approx 2.5 \cdot 10^{-11}$~Pa$^{-1}$.
The experiments therefore show that $\al$ is 10 times larger than $\beta$, and so the parameter $\beta$ can be safely discarded in further calculations.

Next, we use the fact that the total number of electrons $N$ is independent of pressure.
Therefore
\begin{eqnarray}\label{g2}
N=N_P &&= V_P n_P = V_P \int_{\Delta}^{\ee_F^P} g(\ee) d\ee \nonumber \\
&&= CV_P\sqrt{m^*_P}(\ee_F^P- \Delta)^{3/2}(3\ee_F^P+ 2\Delta) \nonumber \\
&&= CV\sqrt{m^*}(\ee_F- \Delta)^{3/2}(3\ee_F+ 2\Delta).
\end{eqnarray}
Throughout further derivation, we shall assume for simplicity that  $\ee_F \gg \Delta$. The above relation gives
after substituting the values of $\ee_F^P$ and $V_P$
\begin{eqnarray}\label{g3}
N_P &&= CV_P\sqrt{m^*_P}(\ee_F^P- \Delta)^{3/2}(3\ee_F^P+ 2\Delta) \nonumber \\
&& \approx CV(1-\beta P)\sqrt{m^*(P)}(1+\al P)^{5/2} \nonumber \\
&&\,\,\, (\ee_F- \Delta)^{3/2}(3\ee_F+ 2\Delta)
\end{eqnarray}
Equations (\ref{g2}) and (\ref{g3}) give the same total carrier number for ambient pressure $P=0$, and a finite pressure, only if the effective mass is equal to:
\begin{equation}\label{g4}
 m^*(P) = \frac{m^*}{(1+\al P)^5(1-\beta P)^2} \approx {m^*}\left[ 1 - P(5\al -2\beta) \right]
\end{equation}
The above development of $m^*$ is valid because $\alpha$ and $\beta$ are very small with respect to the applied pressure.
The mass decreases with pressure as expected, under the condition that $\alpha > 2\beta/5$. 
This condition is clearly satisfied, as seen above.

To summarize this part, we assume that $m^*$ changes under pressure more than any other model parameter, which is logical seeing the layered structure of the crystal.  It is possible to extract the parameter $\alpha$ in two different ways, from $T^*$ and $\ee_F$, and they  agree quite well. Finally, $\beta$ is much smaller than $\alpha$.

\subsection{Calculation of the $dc$ resistivity anisotropy}

The intraband conductivity tensor is defined in the direction of principal axes $\nu=x,y,z$.
\begin{equation}\label{vod1}
 \s_{\nu\nu}(\om) =  \frac{ie^2}{m_e}  \frac{n_{\nu\nu}}{\om + i\Gamma}
\end{equation}
with the effective concentration of charge carriers $n_{\nu\nu}$ and the relaxation constant $\Gamma$. 
The $dc$ resistivity is then $\vo_{xx} =  1/\s_{xx}(0) $.
The effective number of charge carriers $ n_{\nu\nu} $ is defined and calculated in the Appendix. It is the only relevant parameter in determining the resistivity ratios, as seen from the expression (\ref{vod1}). The out-of-plane resistivity anisotropy $\vo_a/\vo_b$, based on our model, is equal to:
\begin{eqnarray}\label{vod12}
\frac{\vo_{xx}}{\vo_{zz}} = \frac{n_{zz}}{n_{xx}}   \approx \frac{3}{4}\frac{\ee_F}{m^* v_x^2}
 = 1/500.
\end{eqnarray}
The in-plane anisotropy $\vo_c/\vo_a$ is:
\begin{eqnarray}\label{vod13}
\frac{\vo_{yy}}{\vo_{xx}} = \frac{n_{xx}}{n_{yy}} = \frac{v_{x}^2}{v_{y}^2}   \approx 2.2.
\end{eqnarray}

While the experimental in-plane anisotropy, $\vo_c/\vo_a \sim 2$, agrees very well with the model result, the out-of plane anisotropy is experimentally six times smaller.
This may be related to the questionable assumption that the relaxation coefficient in the intraband conductivity is the same in all the three directions, despite a highly anisotropic Fermi surface.
The Fermi surface of such a doped system is an elongated ellipsoid in which the short axes ($x, y$) are similar in length and determined by velocity ratios. In the $z$ direction, the parabolic dispersion covers $\sim70$\% of the Brillouin zone. 

Perhaps more relevant is the fact that the out-of-plane to in-plane resistivity ratio in Eq.~(\ref{vod12}) is highly susceptible to the changes of velocity $v_x$, which may not be known very precisely. If this velocity changes by a factor of 2, the ratio in Eq.~(\ref{vod12}) would give the experimentally observed value in Fig.~\ref{Fig4}b.

\subsection{Pressure and temperature effects on the resistivity}

Finally, we consider the variation of the resistivity at low pressures, below 2 GPa, and at low temperatures. In a metal at a finite temperature, the deviation of the chemical potential $\mu$ from the Fermi level $\ee_F$ is a function of the density of states Eq.~(\ref{t1}), in the leading order of $T$. It is given by:
\begin{equation} \label{t2}
 \mu \approx   \ee_F\left(1-{\tau^2}/{\ee_F^2}\right),
\end{equation}
where $\tau \approx \pi k_BT/2$. 
Analogously, for a finite pressure $P$, the chemical potential Eq.~(\ref{t2}) depends on $\ee_F^P$ as:
\begin{equation}\label{t22}
\mu_P = \ee_F^P \left( 1-{\tau^2}/{(\ee_F^P)^2} \right).
 \end{equation}
Finite values of pressure and temperature $(P,T)$ alter the effective concentration Eq.~(\ref{vod9}) in a trivial way, see Appendix:
\begin{eqnarray}\label{t4}
 n_{xx}(P,T) \approx  n_{xx}(0,0) \left(1-\frac{3}{2} \frac{\tau^2}{\ee_F^2(1+\al P)^2}\right) (1-\al P) \nonumber \\
\end{eqnarray}
This results in the $T^2$ behavior of the resistivity at low temperatures:
\begin{eqnarray}\label{t5}
 \rho_x (P,T ) &=&  \rho_x(0,0)(1+\al P) + A_PT^2
\end{eqnarray}
with a constant $A_P = 0.66 \times 10^{-9}(1-\al P)\, \rm{\Omega m /K^2}$. For our sample A, the calculated prefactor $A_P$ accounts for about 60\% of the experimental value. This is a surprisingly good agreement, considering that there may at the same time be other scattering mechanisms which also give a $T^2$ resistivity dependence.

The full expression for the resistivity, Eq.~(\ref{t5}), can be evaluated numerically, and the results are shown in Fig.~\ref{Fig5}c for three different pressures. We note that the temperature of the resistivity peak, $T^*$, shifts linearly in temperature under pressure, in agreement with our experimental data.
Moreover, our model predicts an increase of the zero temperature resistivity, $\varrho_0$, under pressure, and a decrease of $A_P$. We observe both of these effects, in particular in the sample B which does not have low-temperature upturns.

\section{Conclusion}
Based on our high-pressure transport and infrared experiments, ZrTe$_5$ is confirmed to be very sensitive to hydrostatic pressure.
In the high-pressure  resistivity, we observe a strong increase of the peak temperature $T^*$ even for fairly low pressures. In the high-pressure optical transmission, one can follow a linear increase of the absorption onset, again at low pressures. Both of these observations point to an increase in the Fermi energy as the main effect of low pressures. 
Within our two-band, low-energy effective model, these experimental observations can be well explained by the decrease of the effective mass under pressure.
Finally, based on the effective model, we derive expressions for the resistivity in the low-temperature and low-pressure regime.

\section{Acknowledgements}
The authors acknowledge illuminating discussions with L. Forr\'o and E. Svanidze, experimental help by F. Capitani and F. Borondics, as well as kind help by N. Miller.
We also thank A. Crepaldi for his generous help with samples, and for extensive discussions.
A.~A. acknowledges funding from the  Swiss National Science Foundation through project PP00P2\_170544.
This work has been supported by the ANR DIRAC3D. We acknowledge the support of LNCMI-CNRS, a member of the European Magnetic Field Laboratory (EMFL).
A part of this work was done at the SMIS beamline of Synchrotron Soleil, Proposal 20170565.
Work at BNL was supported by the U.S. Department of Energy, Office of Basic Energy Sciences, Division of Materials Sciences and Engineering under Contract No. DE-SC0012704.

\begin{widetext}

\section{Appendix}

\subsection{Ratio of the $dc$ resistivity $\vo_{zz}/\vo_{xx}$}

The intraband conductivity tensor is defined by:
\begin{equation}\label{vod1a}
 \s_{xx}(\om) =  \frac{ie^2}{m_e}  \frac{n_{xx}}{\om + i\Gamma}
\end{equation}
with the effective concentration of charge carriers $n_{xx}$ and the relaxation constant $\Gamma$. 
The $dc$ resistivity is then: 
\begin{equation}\label{vod2}
\vo_{xx} =  1/\s_{xx}(0) =  \frac{m_e\Gamma}{e^2n_{xx}} 
\end{equation}
The effective number of charge carriers $n_{\al \al}$ is given by: 
\begin{equation}\label{vod3}
 n_{\al \al} = - \frac{1}{V} \sum_{ \kk \s}m_e v_{\al \kk}^2 \frac{\partial f_{\kk}}{ \partial \ee_{\kk}},
\end{equation}
where $v_{\al \kk} = (1/\hbar){\partial \ee_{\kk}}/{ \partial k_{\al}}$. At low temperatures ($T \approx 0$), we have ${\partial f_{\kk}}/{ \partial \ee_{\kk}} = -\delta(\ee_F - \ee_{\kk})$,
and for $\al = x,z$ we have
\begin{equation}\label{vod4}
 {\partial \ee_{\kk}}/{ \partial k_{x}} = \frac{(\hbar v_x)^2k_x}{\sqrt{\hbar^2(v_xk_x)^2 + \hbar^2(v_yk_y)^2 + (\Delta + ck_z^2 )^2 }} = \frac{(\hbar v_x)^2k_x}{\ee_{\kk}}
\end{equation}
and
\begin{equation}\label{vod5}
 {\partial \ee_{\kk}}/{ \partial k_{z}} = \frac{(\Delta + ck_z^2)2ck_z}{\sqrt{\hbar^2(v_xk_x)^2 + \hbar^2(v_yk_y)^2 + (\Delta + ck_z^2 )^2 }} = \frac{(\Delta + ck_z^2)2ck_z}{\ee_{\kk}}
\end{equation}
Next, we calculate the effective number of charge carriers in $x$ and $z$ direction. By introducing new dimensionless variables $v_x\hbar k_x = x$, $v_y\hbar k_y = y$ and $k_z\sqrt{c} = z$, the integral  Eq.~(\ref{vod3}) becomes
 
 \begin{eqnarray}\label{vod6}
 n_{xx} = \frac{1}{(2\pi)^3} \frac{2m_e}{\hbar^2} \frac{v_x}{v_y}\frac{1}{\sqrt{c}} \frac{1}{\ee_F^2} \iiint x^2 {\delta \left(\ee_F - \sqrt{x^2 + y^2 + (\D + z^2)^2}\right)} dxdydz
\end{eqnarray}
Changing from Cartesian to cylindrical coordinates by introducing $\vo^2 =x^2 + y^2 $, Eq.~(\ref{vod3}) becomes: 
\begin{eqnarray}\label{vod7}
 n_{xx} =\frac{1}{(2\pi)^3} \frac{2m_e}{\hbar^2} \frac{v_x}{v_y}\frac{1}{\sqrt{c}} \frac{1}{\ee_F^2} \int_0^{\infty} d\vo \int_0^{2\pi} d\varphi \int_{-\infty}^{\infty}dz \,
 \vo^3 {\rm{cos}}^2\varphi {\delta \left(\ee_F - \sqrt{\vo^2 + (\D + z^2)^2}\right)}
\end{eqnarray}
We note that there are two zeros of $z$ argument within the $\delta$ function  $z_0 = \pm \sqrt{\sqrt{\ee_F^2 - \vo^2} - \D}$.
This gives an extra factor of 2, and the integral reduces to 
\begin{eqnarray}\label{vod8}
 n_{xx} =\frac{1}{(2\pi)^3} \frac{2m_e}{\hbar^2} \frac{v_x}{v_y}\frac{1}{\sqrt{c}}  \frac{\pi}{\ee_F} \int_0^{\sqrt{\ee_F^2 - \D^2}}\!\! \frac{\vo^3 d\vo}{\sqrt{\ee_F^2 - \vo^2}\sqrt{\sqrt{\ee_F^2 - \vo^2} - \D}}  
\end{eqnarray}
which can be easily evaluated, giving the final result: 
\begin{eqnarray}\label{vod9}
 n_{xx} =\frac{1}{(2\pi)^2} \frac{8}{15}\frac{m_e}{\hbar^3} \frac{v_x}{v_y}\sqrt{2m^*} \frac{\sqrt{\ee_F-\Delta}}{\ee_F}(3\ee_F^2 - \Delta \ee_F - 2\Delta^2).
\end{eqnarray}
Similarly we can calculate $n_{zz}$. The integration is a bit more complex due to the extra term in the numerator of Eq.~(\ref{vod5}). Here we only give the result, since the derivation is similar:
\begin{eqnarray}\label{vod10}
 n_{zz} =\frac{1}{(2\pi)^2} \frac{2m_e}{\hbar^3}\frac{2}{105} \frac{1}{v_xv_y}\frac{4}{\sqrt{2m^*}} \frac{\sqrt{\ee_F-\Delta}}{\ee_F}(15\ee_F^3 - 4\Delta^2 \ee_F-3\Delta\ee_F^2 - 8\Delta^3)
\end{eqnarray}
The ratio of the two quantities can be written in the form  
\begin{eqnarray}\label{vod11}
\frac{n_{zz}}{n_{xx}} = \frac{15}{105}\frac{1}{m^* v_x^2} \left( \frac{15\ee_F^2}{3\ee_F+2\Delta} +4\Delta \right) \approx \frac{3}{4}\frac{\ee_F}{m^* v_x^2}.
\end{eqnarray}
Finally, by inspecting the expression (\ref{vod2}) we can determine 
\begin{eqnarray}\label{vod12a}
\frac{\vo_{xx}}{\vo_{zz}} = \frac{n_{zz}}{n_{xx}}   \approx 1/500,
\end{eqnarray}
while from Eq.~(\ref{vod9}) we can conclude
\begin{eqnarray}\label{vod13}
\frac{\vo_{yy}}{\vo_{xx}} = \frac{v_{x}^2}{v_{y}^2}   \approx 2.2
\end{eqnarray}

\subsection{Effective carrier concentration $n_{xx}(P,T)$}

In metals at finite temperatures the deviation of the chemical potential $\mu$ from $\ee_F$ is 
\begin{equation} \label{t1}
 \mu \approx \ee_F \left[ 1-\frac{\pi^2}{6}\frac{(k_BT)^2}{\ee_F} \frac{1}{g(\ee)} \left.\frac{\partial g(\ee)}{ \partial \ee} \right|_{\ee_F}  \right]
\end{equation}
or inserting Eq.~(\ref{g2}) in Eq.~(\ref{t1})
\begin{equation} \label{t2}
 \mu \approx \ee_F \left[ 1-\frac{\pi^2 (k_BT)^2}{12 \ee_F^{2}} \frac{3\ee_F - 2\D}{\ee_F - \D}   \right] \approx   \ee_F\left(1-\frac{\tau^2}{\ee_F^2}\right),
\end{equation}
where $\tau = \pi k_BT/2$. Assuming a finite pressure $P$, we have: 
\begin{equation}\label{t22}
 \mu \to \mu_P = \ee_F^P \left( 1-\frac{\tau^2}{(\ee_F^P)^2} \right)=\ee_F(1+\al P)\left( 1-\frac{\tau^2}{\ee_F^2(1+\al P)^2} \right) = \ee_F(1+\al P)(1-\eta)
 \end{equation}
where we have $\eta \ll 1$. Finite values of $(P,T)$ alter the effective concentration Eq.~(\ref{vod9}) in a trivial way
\begin{equation}\label{t3}
 n_{xx}(P,T) =\frac{1}{(2\pi)^2} \frac{8}{15}\frac{m_e}{\hbar^3} \frac{v_x}{v_y}\sqrt{2m^*_P} \frac{\sqrt{\mu_P-\Delta}}{\mu_P}(3\mu_P^2 - \Delta \mu_P - 2\Delta^2).
\end{equation}
Inserting Eqs.~(\ref{g4}) and (\ref{t2}) into the above relation, and assuming that $\ee_F \gg \Delta$, we have:
\begin{eqnarray}\label{t4}
 n_{xx}(P,T) && \approx \frac{1}{(2\pi)^2} \frac{8}{15}\frac{m_e}{\hbar^3} \frac{v_x}{v_y}\sqrt{2m^*} \frac{\sqrt{\ee_F-\Delta}}{\ee_F}(3\ee_F^2 - \Delta \ee_F - 2\Delta^2) 
 \frac{1-2\eta}{\sqrt{1-\eta}} \frac{1+2\al P}{(1+\al P)^3}
 \nonumber \\
   && \approx  n_{xx}(0,0) (1-3\eta/2)(1-\al P)
\end{eqnarray}
Finally, changing the variable $\eta$ to its explicit form,  Eq.~(\ref{t22}), the resistivity behaves like:
\begin{eqnarray}\label{t5}
 \rho_x (P,T )= 1/ \s_x(P,T) &=&  \frac{m_e\Gamma}{e^2n_{xx}(P,T)} \approx   \frac{m_e\Gamma}{e^2n_{xx}(0,0)}(1+3\eta/2)(1+\al P) \nonumber \\
 &=&  \rho_x(0,0)(1+\al P) + A_PT^2
\end{eqnarray}
with the constant $A_P = 0.66 \times 10^{-9}(1-\al P)\, \rm{\Omega m /K^2}$.

\end{widetext}

\bibliography{ZrTe5}

\end{document}